\documentclass[twocolumn,aps,superscriptaddress]{revtex4}
\usepackage{graphicx}
\usepackage{float}
\usepackage{dcolumn}
\usepackage{bm}
\usepackage{color}
\usepackage{amsmath}
\usepackage{amssymb}
\usepackage{amsfonts}
\usepackage{esint}
\usepackage{times}
\usepackage{xcolor}
\usepackage{phaistos}
\usepackage{braket}
\usepackage{comment}

\usepackage{pdfpages}

\usepackage[colorlinks,linkcolor=blue,anchorcolor=blue,citecolor=blue]{hyperref}

\begin{document}

\title{Combating fluctuations in relaxation times of fixed-frequency transmon qubits \\ with microwave-dressed states}
\author{Peng Zhao}
\email{shangniguo@sina.com}
\affiliation{Beijing Academy of Quantum Information Sciences, Beijing 100193, China}
\author{Teng Ma}
\email{mateng@baqis.ac.cn}
\affiliation{Beijing Academy of Quantum Information Sciences, Beijing 100193, China}
\author{Yirong Jin}
\email{jinyr@baqis.ac.cn}
\affiliation{Beijing Academy of Quantum Information Sciences, Beijing 100193, China}
\author{Haifeng Yu}
\affiliation{Beijing Academy of Quantum Information Sciences, Beijing 100193, China}

\date{\today}

\begin{abstract}
With the long coherence time, the fixed-frequency transmon qubit is a promising qubit modality
for quantum computing. Currently, diverse qubit architectures that utilize
fixed-frequency transmon qubits have been demonstrated with
high-fidelity gate performance. Nevertheless, the relaxation times of transmon qubits
can have large temporal fluctuations, causing instabilities in gate performance. The
fluctuations are often believed to be caused by nearly on-resonance
couplings with sparse two-level-system (TLS) defects. To mitigate their impact on qubit coherence and gate performance, one direct
approach is to tune the qubits away from these TLSs. In this work, to combat the potential TLS-induced performance fluctuations
in a tunable-bus architecture unitizing fixed-frequency transmon qubits, we explore the
possibility of using an off-resonance microwave drive to effectively tuning the qubit frequency
through the ac-Stark shift while implementing universal gate operations on the
microwave-dressed qubit. We show that the qubit frequency can be tuned up to $20\,\rm MHz$ through
the ac-stark shift while keeping minimal impacts on the qubit control. Besides passive approaches
that aim to remove these TLSs through more careful treatments of device fabrications, this
work may offer an active approach towards mitigating the TLS-induced performance fluctuations
in fixed-frequency transmon qubit devices.

\end{abstract}

\maketitle

\section{Introduction}

With the advantages of long coherence times \cite{Kjaergaard2020,Place2021,Wang2022,Gordon2022} and
simplified demands on the electronic control circuits, fixed-frequency transmon qubit \cite{Koch2007} has been
demonstrated as a leading qubit modality for quantum computing \cite{Kjaergaard2020}. This can be
partially manifested by the progress that diverse qubit architectures, which utilize
fixed-frequency transmon qubits, have been demonstrated with high-fidelity gate
performance, such as all-microwave controlled qubit architecture with fixed inter-qubit
coupling \cite{Chow2012,Poletto2012,Chow2013,Paik2016,
Premaratne2019,Krinner2020,Mitchell2021,Wei2021}, qubit architecture with tunable bus \cite{McKay2016}
or tunable coupler \cite{Collodo2020,Xu2020}, and qubit architecture combining both fixed-frequency qubits and
frequency-tunable qubits \cite{Hong2020}. However, just the same as
frequency-tunable transmon qubits \cite{Klimov2018}, the relaxation times of fixed-frequency
transmon qubits can show large temporal fluctuations \cite{Burnett2019,Schlor2019,Carroll2021}.
Since the current gate performance approaches the qubit coherence
limit \cite{Kjaergaard2020}, the fluctuations can lead to
prominent performance instabilities in transmon qubit devices \cite{Hong2020,Kim2021}.

For transmon qubit devices, the fluctuations in relaxation times
are often believed to be attributed to nearly on-resonance couplings with sparse
two-level-system (TLS) defects \cite{Klimov2018,Burnett2019,Schlor2019,Carroll2021}.
The TLS defect can act as environmental noise coupled to the qubit with the noise spectral
density peaked around its frequency. As the frequency of the TLS defect can
have temporal fluctuations due to its couplings to thermally fluctuating defects (i.e., low-frequency TLSs with
frequencies less than $k_{B}T$, where $k_{B}$ is the Boltzmann constant and $T$ is the working
temperature of the qubit devices, typically, 10 mK), this could explain the fluctuations in
relaxation times of transmon qubits \cite{Burnett2014,Muller2015,Faoro2015,Bejanin2021}.
Meanwhile, in conditions where qubit relaxations contain non-negligible
contributions from quasiparticles, the variations of qubit relaxation times
could also be attributed to the fluctuation in the quasiparticle density near the qubit
junctions \cite{Vool2014,Gustavsson2016,Yan2016} or the fluctuations in quasiparticle dissipation
channels \cite{Li2021,Zhang2021}. Nevertheless, for well-shielded transmon qubit devices, transmon
qubits have been shown not yet limited by losses related to quasiparticles \cite{Gordon2022}.
Additionally, a recent result also suggests that quasiparticles trapped in shallow subgap
states can also behave similarly to TLS \cite{Graaf2020}.

Generally, the existence of performance fluctuations suggests that, to maintain a reliable and
high-fidelity gate performance, rapid and frequent qubit characterizations are
needed \cite{Kelly2018}. Moreover, besides being frequently re-characterized for detecting and
tracking these fluctuations, one has also to actively mitigate the detrimental impact on
gate performance. For the state-of-the-art transmon qubit devices, the TLS defects that
are coupled strongly or nearly on-resonantly to qubits are generally
sparse \cite{Klimov2018,Carroll2021}. Thus, to mitigate the TLS-induced performance
fluctuations, the direct and active approach is to tune the qubit away
from the dominant TLS defect, and then a re-calibration of gate operations should be
employed to find new optimal control parameters \cite{Kelly2018}. We note that ultimately, passive
approaches that aim to remove the detrimental TLS defects through more advanced
fabrication technology, should work successfully, but currently, the exact nature of the TLS defect is
still unknown \cite{Muller2019} and this active approach may be a more practical
solution.

For fixed-frequency transmon qubits, to mitigate TLS-induced performance fluctuations,
one can effectively tune its frequency through the off-resonance drive induced ac-Stark shift \cite{Tuorila2010,Schneider2018,Oelsner2020,Carroll2021}, and choose the microwave-dressed qubit states
as the basis states of a new qubit, i.e., the dressed
qubit \cite{Cohen-Tannoudji1998,Rigetti2005,Liu2006,Wilson2007,Guo2018,Wei2021}. In this way,
gate operations should be implemented on this microwave-dressed basis. However,
while the ac-Stark shift has been recognized as an effective tool for tuning qubit frequency, its
compatibility with the gate operations on superconducting qubits is less
studied \cite{Liu2006,Guo2018,Wei2021}. In this work,
to combat the potential TLS-induced performance fluctuations in a tunable-bus
architecture unitizing fixed-frequency transmon qubits \cite{Zhao2021}, we examine the
possibility of tuning the qubit away from the dominant TLS defect through the ac-Stark shift
while implementing universal qubit control, including gate operations, qubit
initialization, and readout, on the microwave-dressed qubit. We show that although parasitic
interactions induced by the stark drive exist, one can still mitigate their detrimental impacts on qubit
control. Thus, using the stark drive, the qubit frequency can be tuned up to $20\,\rm MHz$ while keeping
minimal impacts on the qubit control.

The rest of the paper is organized as follows. In Sec.~\ref{SecII}, we give brief descriptions of the
microwave-dressed qubit, showing how to mitigate the detrimental effect from
a dominated TLS defect through the ac-Stark shift. In Sec.~\ref{SecIII}, in a tunable-bus architecture,
we examine the possibility of implementing universal qubit control, including
gate operations, qubit initialization, and readout, for the microwave-dressed qubit.
In Sec.~\ref{SecIV}, we give discussions on the feasibility of the
proposed scheme for mitigating TLS-induced performance fluctuations.
Finally, we give conclusions of our investigation in Sec.~\ref{SecV}.

\begin{figure}[tbp]
\begin{center}
\includegraphics[keepaspectratio=true,width=\columnwidth]{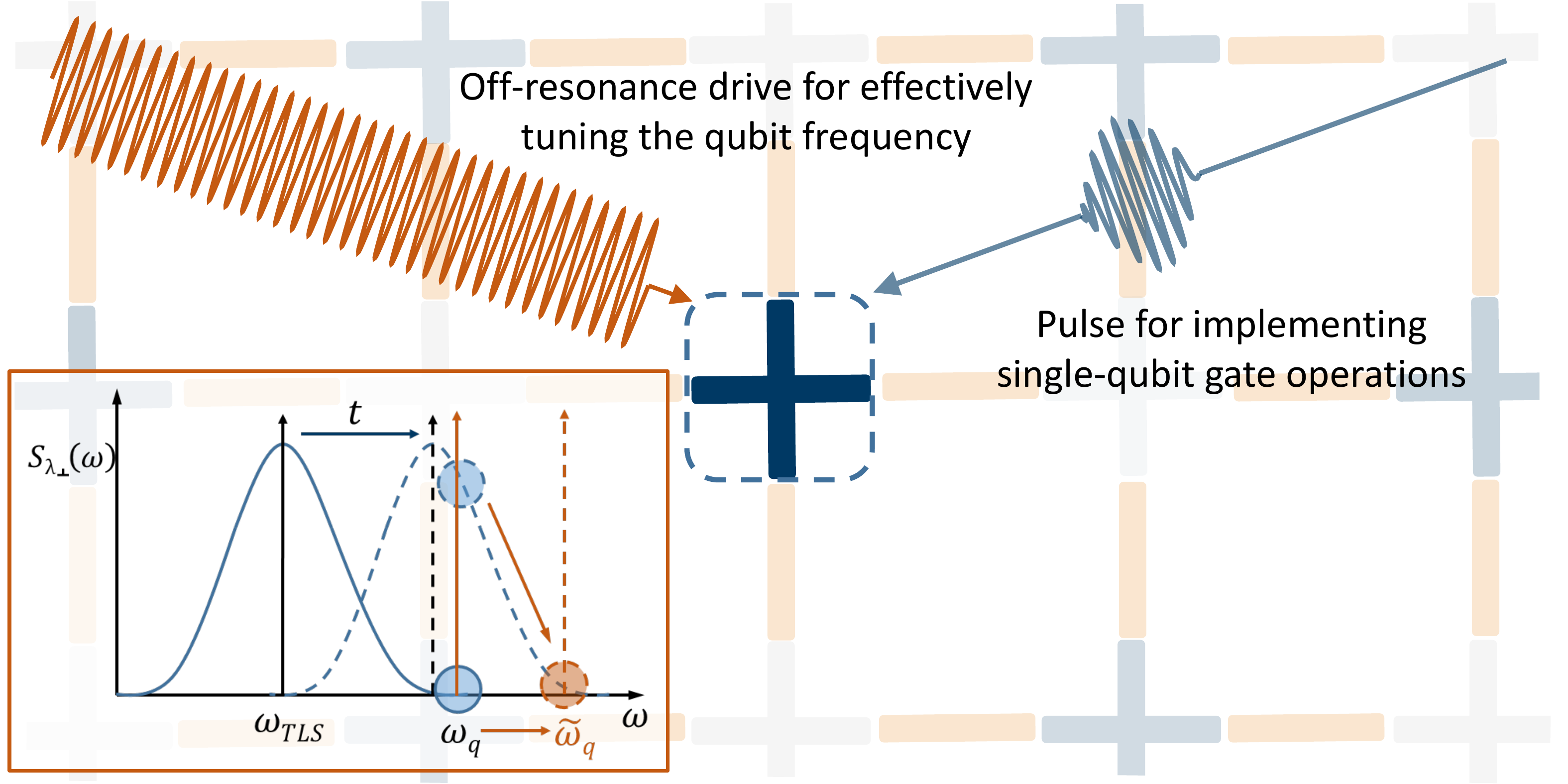}
\end{center}
\caption{Illustration of mitigating TLS-induced relaxation fluctuations of a qubit through the
off-resonance drive induced ac-Stark shift. The inset shows how the fluctuation
in qubit relaxation times occurs due to its coupling to a dominant TLS for which its frequency
can have temporal fluctuation, and how to mitigate the relaxation fluctuations through tuning the
qubit away from the TLS, i.e., tuning from $\omega_{q}$ (solid orange (light gray) arrow)
to $\tilde{\omega}_{q}$ (dashed orange (light gray) arrow). Blue solid curve denotes the
noise spectral density peaked at the TLS's frequency $\omega_{TLS}$ (solid black
arrow), blue dashed line for the case when fluctuations in the TLS's frequency
occurs (dashed black arrow) due to its couplings to thermally fluctuating
defects (low-frequency TLS defects).}
\label{fig1}
\end{figure}

\section{Combating TLS-induced fluctuations with Microwave-dressed states}\label{SecII}

Here, firstly, for easy reference and to set the notation, we briefly review some basic
properties of the qubit subjected to an off-resonance stark drive, including the ac-Stark shift
and the decoherence of the qubit in the presence of the stark drive. Accordingly, in the following
discussion, we call the qubit subjected to a stark drive as microwave-dressed qubit or
dressed qubit, and call the undriven qubit as the bare qubit. Then, we illustrate the
mechanism of mitigating the detrimental effect from a dominated TLS defect through the ac-Stark shift.

\subsection{ac-Stark shift}\label{SecIIA}

Here, as shown in Fig.~\ref{fig1}, we consider a transmon qubit driven by an always-on
off-resonance microwave drive. For illustration purposes only, the qubit is treated as
an ideal two-level system. Thus, the Hamiltonian of this driven qubit system can be written as
$H=\omega_{q}\sigma_{z}/2+\Omega_{s}\cos(\omega_{s}t)\sigma_{x}$ (hereinafter, we set $\hbar=1$),
where $\omega_{q}$ denotes the bare qubit frequency, $\omega_{s}$ and $\Omega_{s}$ represent the
frequency and the amplitude of the drive (hereafter called stark drive) which is introduced to
induce an ac-Stark shift for the qubit. After applying the rotating wave
approximation (RWA), and moving into the rotating frame with respect to the stark drive,
the Hamiltonian becomes $H_{\rm r}=\Delta_{s}\sigma_{z}/2+\Omega_{s}\sigma_{x}/2$,
where $\Delta_{s}=\omega_{q}-\omega_{s}$ denotes the detuning of the qubit from the stark
drive frequency. Finally, considering the unitary transformation $U_{1}={\rm exp}(-i\theta\sigma_{y}/2)$
with $\theta=\arctan(\Omega_{s}/\Delta_{s})$,
$H_{r}$ can be diagonalized and expressed as $H_{\rm eff}=\Delta Z/2$ with $\Delta=\sqrt{\Delta_{s}^{2}+\Omega_{s}^{2}}$. Here, $Z\equiv\cos{\theta}\sigma_{z}+\sin{\theta}\sigma_{x}$
denotes the Pauli operator defined in the dressed basis (i.e., the eigenstates of $H_{r}$)
\begin{equation}
\begin{aligned}\label{eq1}
\{&|1\rangle\equiv\sin\frac{\theta}{2}|g\rangle+\cos\frac{\theta}{2}|e\rangle,
\\&|0\rangle\equiv\cos\frac{\theta}{2}|g\rangle-\sin\frac{\theta}{2}|e\rangle\},
\end{aligned}
\end{equation}

According to the above discussion, for the microwave-driven qubit, the ac-Stark
shift can be expressed as $\delta\omega=\Delta-\Delta_{s}\approx \Omega_{s}^{2}/(2\Delta_{s})$.
However, since the transmon qubit is naturally a multilevel
system with a weak anharmonicity \cite{Koch2007}, its higher energy levels can give
non-negligible contributions on the stark shift $\delta\omega$. Thus, modeling the transmon qubit as
an anharmonic oscillator with the anharmonicity $\eta$ and taking the higher energy levels, especially the
second excited state $|f\rangle$, into considerations, the Ac-stark shift \cite{Schneider2018,Carroll2021}
can be approximated as (the second-order perturbation result)
\begin{equation}
\begin{aligned}\label{eq2}
\delta\omega\approx \frac{\Omega_{s}^{2}}{2\Delta_{s}}-\frac{(\sqrt{2}\Omega_{s})^{2}}{4(\Delta_{s}+\eta)}
=\frac{\eta\Omega_{s}^{2}}{2\Delta_{s}(\Delta_{s}+\eta)}.
\end{aligned}
\end{equation}

\subsection{Decoherence of the microwave-dressed qubit}\label{SecIIB}

We now discuss the environment-induced decoherence of the microwave-dressed qubit.
The environment noise coupled to the qubit can be described as
$H_{\delta\lambda}=\delta\lambda_{z}\sigma_{z}/2+\delta\lambda_{\bot}\sigma_{\bot}/2$,
where $\sigma_{\bot}$ represents the transverse Pauli component, $\delta\lambda_{z}$
and $\delta\lambda_{\bot}$ denote the fluctuations in qubit parameters caused by
the environment noise transversally and longitudinally coupled to
the qubit, respectively.

Here, we consider that $S_{\lambda}(\omega)\equiv1/(2\pi)\int dt\langle\delta\lambda(0)\delta\lambda(t)\rangle e^{-i\omega t}$
denotes the quantum noise spectral density associated with parameter $\lambda$.
Taking into account that the transmon qubit devices generally work at very low
temperatures ($\omega_{q}\gg k_{B}T$), we neglect the absorption process involving $S_{\lambda_{\bot}}(-\omega)$,
and only consider the emission process involving $S_{\lambda_{\bot}}(+\omega)$ (here, $\omega\simeq\omega_{q}$) \cite{Ithier2005}.
Thus, the relaxation time $\tilde{T}_{1}$ and the pure dephasing time $\tilde{T}_{\phi}$ of the
microwave-dressed qubit can be expressed as \cite{Ithier2005,Jing2014}
\begin{equation}
\begin{aligned}\label{eq3}
&\frac{1}{\tilde{T}_{1}}=\pi\left[\frac{1+\cos^{2}\theta}{4}\frac{\tilde{S}_{\lambda_{\bot}}(\omega_{s},\Delta)}{2}+\sin^{2}\theta S_{\lambda_{z}}(\Delta)\right],
\\&\frac{1}{\tilde{T}_{\phi}}=\pi\left[\cos^{2}\theta S_{\lambda_{z}}(0)+\frac{\sin^{2}\theta}{4}S_{\lambda_{\bot}}(\omega_{s})\right],
\end{aligned}
\end{equation}
with $\tilde{S}_{\lambda_{\bot}}(\omega_{s},\Delta)=S_{\lambda_{\bot}}(\omega_{s}+\Delta)+S_{\lambda_{\bot}}(\omega_{s}-\Delta)$.
The above equation shows that due to the stark drive induced state hybridization, the environmental noise,
transversally (longitudinally) coupled to the bare qubit, can contribute to the dephasing (relaxation) of
the microwave-dressed qubit. Moreover, for the qubit relaxation, both the transverse noises at the
sideband frequencies $\omega_{s}+\Delta$ and $\omega_{s}-\Delta$ contribute to the relaxation (depolarization)
of the dressed qubit. This notable feature can be captured
by a qualitative picture, i.e., depending on the parameters (e.g., detuning and magnitude) of
external drives (stark drives), the drives could act as distinct noise spectral
filters, allowing the qubits, i.e., spectrometers, to sense the environment noise
differently in frequency space \cite{Jing2014,Green2013}.

\subsection{Mitigating TLS-induced coherence degradation}\label{SecIIC}

For the state-of-the-art transmon qubit devices, the TLS defects that
are coupled strongly or nearly on-resonantly to qubits are generally
sparse \cite{Klimov2018,Carroll2021}. Therefore, one may reasonably expect that when
prominent degradations in qubit relaxation times occur, the dominated contribution could
be attributed to a single TLS defect that is coupled nearly on-resonantly to the
qubit \cite{Klimov2018,Carroll2021,Schlor2019}. As shown in the inset of Fig.~\ref{fig1}, as the dominated TLS defect
can generally act as an environmental noise with the spectral density $S_{\lambda_{\bot}}(\omega)$
peaked around its frequency $\omega_{TLS}$, pushing the qubit frequency away from
$\omega_{TLS}$, i.e., from $\omega_{q}$ to $\tilde{\omega}_{q}$, can mitigate its
impact on the qubit relaxation time. This mitigation procedure can be further explained
by Eq.~(\ref{eq3}), which gives the expression of the relaxation rate $1/\tilde{T}_{1}$.
Specifically speaking, for the TLS-induced noise peaked at $\omega_{TLS}$, by carefully
choosing the stark drive frequency $\omega_{s}$ and the drive amplitude $\Omega_{s}$,
only the noise at $\omega_{s}\pm\Delta$, in which the amplitude of the noise spectral
density is far less than that at $\omega_{TLS}$, contributes
to the qubit relaxation. Thus, the ac-Stark shift
can be introduced to protect the qubit from the TLS noise peaked at its
frequency.

Note that compared with the bare qubit, there are two additional decoherence
channels for the microwave-dressed qubit: (i) the amplitude-fluctuations,
phase-fluctuations, and frequency-fluctuations of the stark drive can also act as an
additional channel contributing to qubit dephasing or relaxation \cite{Ball2016,Dijk20219,Werninghaus2021,Wei2021};
(ii) For the case of a two-level qubit, the computational basis, i.e., Eq.~(\ref{eq1}),
only involves the $\{|g\rangle, |e\rangle\}$. However, since the transmon qubit is
a multilevel system, the higher energy levels of the transmon qubit, such as $|f\rangle$, are also involved
in the definition of the computational basis for the dressed transmon qubit. This could
give rise to an additional dephasing or relaxation channel \cite{Mitchell2021,Xiong2021}.

\begin{figure*}[tbp]
\begin{center}
\includegraphics[width=18cm,height=9cm]{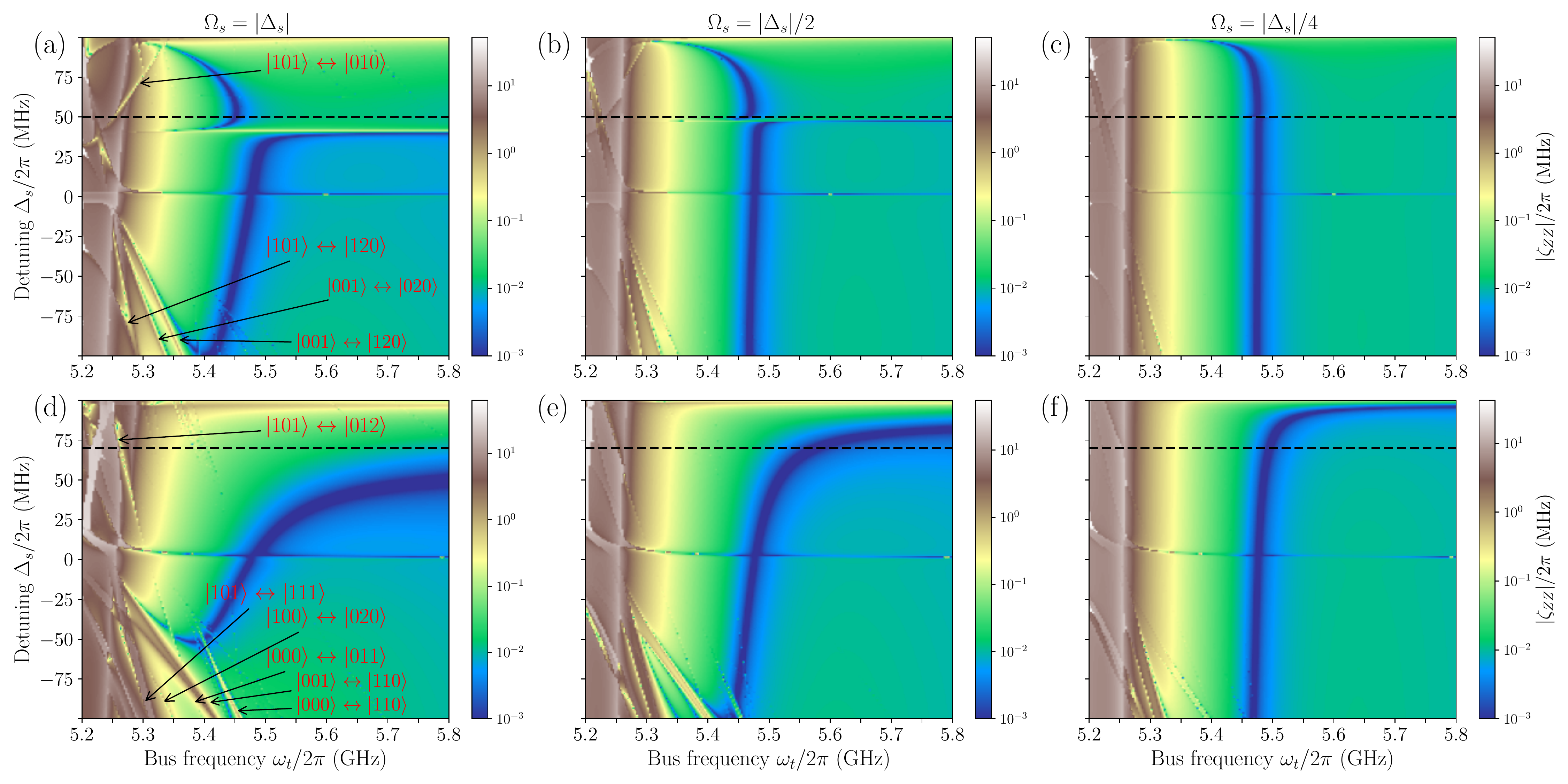}
\end{center}
\caption{$ZZ$ coupling strength $\zeta_{\rm ZZ}$ as a function of the bus frequency ($\omega_{t}$) and
the detuning ($\Delta_{s}$) of qubit from the the applied stark drive frequency. Here, the system parameters are: qubit frequency
$\omega_{1(2)}/2\pi=5.0(5.2)\,\rm GHz$, anharmonicity $\eta_{1}=\eta_{2}=\eta_{t}=\eta\,(\eta/2\pi=-300\,\rm MHz)$, and
qubit-bus coupling $g_{1(2)}=g\,(g/2\pi=25\,\rm MHz)$ (at $\omega_{1(2)}=\omega_{t}=5.5\,\rm {GHz})$.
(a-c) The stark drive is applied to $Q_{1}$, and (d-f) for $Q_{2}$. The amplitude $\Omega_{s}$ of the
stark drive are $|\Delta_{s}|$ in (a,d), $|\Delta_{s}|/2$ in (b,e), and $|\Delta_{s}|/4$ in (c,f). The
black arrows indicate the dips and peaks due to the presence of the stark drive induced parasitic
interactions (red (light gray) texts). These interactions involve multi-excitation transitions, which
are enabled by high-order processes, as indicated by the states involved in the transitions. Thus,
the strengths are generally far smaller than that of the qubit-bus couplings and the drive. Moreover,
the stark drive also act as the leading contribution. This can explain that by decreasing the
drive amplitude $\Omega_{s}$, the dips and peaks slowly
disappear. The horizontal black
dashed lines indicate examples of the stark drive frequencies ($\Delta_{s}$) for $Q_{1}$ and $Q_{2}$ that
are chosen to avoid the detrimental parasitic interactions caused by the applied stark drives.}
\label{fig2}
\end{figure*}

\section{qubit control with dressed-basis in tunable-bus architectures}\label{SecIII}

In this section, we examine the possibility of implementing universal control on the
microwave-dressed transmon qubit, including gate operations, state
initialization, and readout for the dressed qubit. In the following
discussion, we give our analysis on a tunable-bus architecture, where two
fixed-frequency transmon qubits $Q_{1}$ and $Q_{2}$ are coupled via a
tunable bus $Q_{t}$ \cite{Zhao2021}. In this tunable-bus architecture,
the bus can mediate an effective $ZZ$ coupling between the two qubits. One
can turn off the $ZZ$ coupling when implementing single-qubit control, such as single-qubit gate
operations, state initialization, and readout. When turning on the $ZZ$ coupling, two-qubit
controlled-Z (CZ) gates can be realized.

In the presence of a stark drive applied to one of the two fixed-frequency qubits, the system
Hamiltonian can be expressed as (here for illustration purpose only, the stark drive is
applied to $Q_{1}$.)
\begin{equation}
\begin{aligned}\label{eq4}
H=&\sum_{i=1,2,t}\left[\omega_{i}a_{i}^{\dagger}a_{i}+\frac{\eta_{i}}{2}a_{i}^{\dagger}a_{i}^{\dagger}a_{i}a_{i}\right]
\\&+\sum_{k=1,2}\left[g_{k}(a_{t}^{\dagger}a_{k}+a_{t}a_{k}^{\dagger})\right]
\\&+\frac{\Omega_{s}}{2}(a_{1}^{\dagger}e^{-i\omega_{s}t}+a_{1}e^{+i\omega_{s}t}),
\end{aligned}
\end{equation}
where the subscript $i=\{1,2,t\}$ labels $Q_{i}$ with anharmonicity
$\eta_{i}$ and bare mode frequency $\omega_{i}$, $a_{i}\,(a_{i}^{\dagger})$ is
the annihilation (creation) operator for $Q_{i}$, and $g_{k}$ denotes
the coupling strength between the tunable bus $Q_{t}$ and the qubit $Q_{k}$.
As in Sec.~\ref{SecIIA}, in the presence of the stark drive
applied to $Q_{1}$, we can define the microwave-dressed qubit states of $Q_{1}$ as
the basis states for qubit control. In this way, the tunable bus can mediate an
effective $ZZ$ coupling between the bare qubit $Q_{2}$ and the
microwave-dressed qubit $Q_{1}$ \cite{Zhao2021}.

Similar to the procedure given in Sec.~\ref{SecIIA}, after applying the RWA and moving
into the rotating frame with respect to the stark drive, the $ZZ$ coupling strength, which is defined as
$\zeta_{\rm ZZ}\equiv(E_{11}-E_{10})-(E_{01}-E_{00})$, can be obtained by diagonalizing
the system Hamiltonian. Here, $E_{jk}$ denotes eigenenergy of microwave-dressed qubit system
associated with doubly dressed eigenstate $|\tilde{jk}\rangle$ (involving both
state hybridizations from the stark drive and the qubit-bus coupling), which is adiabatically
connected to the bare state $|j0k\rangle$ (hereafter, notation $|Q_{1}\,Q_{t}\,Q_{2}\rangle$ is used,
denoting the system state). Figures ~\ref{fig2}(a-c) show the strength of
the $ZZ$ coupling between $Q_{2}$ and dressed $Q_{1}$ as a function of the bus
frequency $\omega_{t}$ and the detuning $\Delta_{s}$ (here denoting the detuning of $Q_{1}$
from the stark drive). The used system parameters are: qubit frequency
$\omega_{1(2)}/2\pi=5.0(5.2)\,\rm GHz$,
anharmonicity $\eta_{1}=\eta_{2}=\eta_{t}=\eta$ with $\eta/2\pi=-300\,\rm MHz$, and
qubit-bus coupling $g_{1(2)}=g\sqrt{\omega_{1(2)}\omega_{t}/\omega_{\rm ref}^{2}}$
with $g/2\pi=25\,\rm MHz$ and $\omega_{\rm ref}/2\pi=5.5\,\rm {GHz}$ \cite{Barends2019}. In Figs.~\ref{fig2}(a),~\ref{fig2}(b), and~\ref{fig2}(c),
the stark drive amplitudes $\Omega_{s}$ are $|\Delta_{s}|$, $|\Delta_{s}|/2$, and $|\Delta_{s}|/4$,
respectively. In Figs.~\ref{fig2}(d-f), we also show the results for the stark drive
applied to $Q_{2}$, showing the $ZZ$ coupling between the bare qubit $Q_{1}$ and the
microwave-dressed qubit $Q_{2}$.

As shown in Figs.~\ref{fig2}, compared with the result in Ref.\cite{Zhao2021}, the presence
of the stark drive causes two prominent features:

(i) Both the zero-$ZZ$ point and the interaction point for resonance coupling
between $|101\rangle$ and $|020\rangle$, which are used for implementing
CZ gates \cite{Zhao2021}, are shifted. In addition, in some scenarios,
the zero-$ZZ$ point can disappear. This is to be expected, as the off-resonance
stark drive can contribute an additional $ZZ$ coupling due to different
ac-Stark shifts of the computational states \cite{Noguchi2020,Xu2021,Xiong2021,Mitchell2021,Wei2021,Ni2021}.

(ii) There are several dips and peaks in the $ZZ$ coupling. After examining
the system spectrum, we find that these dips and peaks result from the resonance
interactions involving the qubits and the bus, as marked in Fig.~\ref{fig2}. Moreover,
the stark drive induced transitions also participate in the processes that enable
these resonance interactions. This can explain that by decreasing the drive amplitude
$\Omega_{s}$, e.g., from Fig.~\ref{fig2}(a) to Fig.~\ref{fig2}(c), these dips and
peaks slowly disappear. Since these interactions are enabled by high-order
processes, the coupling strengths are generally far smaller than that of the qubit-bus couplings
and the off-resonance drive, and the energy gaps of their associated anti-crossings
typically range from sub-$\rm MHz$ to a few $\rm MHz$. Similar to the result discussed
in Ref.\cite{Zhao2021}, for implementing fast diabatic CZ gates \cite{Martinis2014}, the
presence of these parasitic interactions will give rise to a trade-off between
the error resulting from the desired interaction involving $|101\rangle$ and $|020\rangle$ and
the error from these parasitic resonance interactions with tiny energy gaps. Generally
speaking, during the CZ gate operations, a slow gate-speed is better for mitigating
the leakage from $|101\rangle$ to $|020\rangle$ (or leakage involving other interactions
with larger coupling strengths, such as the bus-qubit interactions, which can potentially
cause the leakage from the qubits to the bus). However, to mitigate the leakage
error from these parasitic interactions with tiny anti-crossings, short-time
gates are better.

\begin{figure}[tbp]
\begin{center}
\includegraphics[keepaspectratio=true,width=\columnwidth]{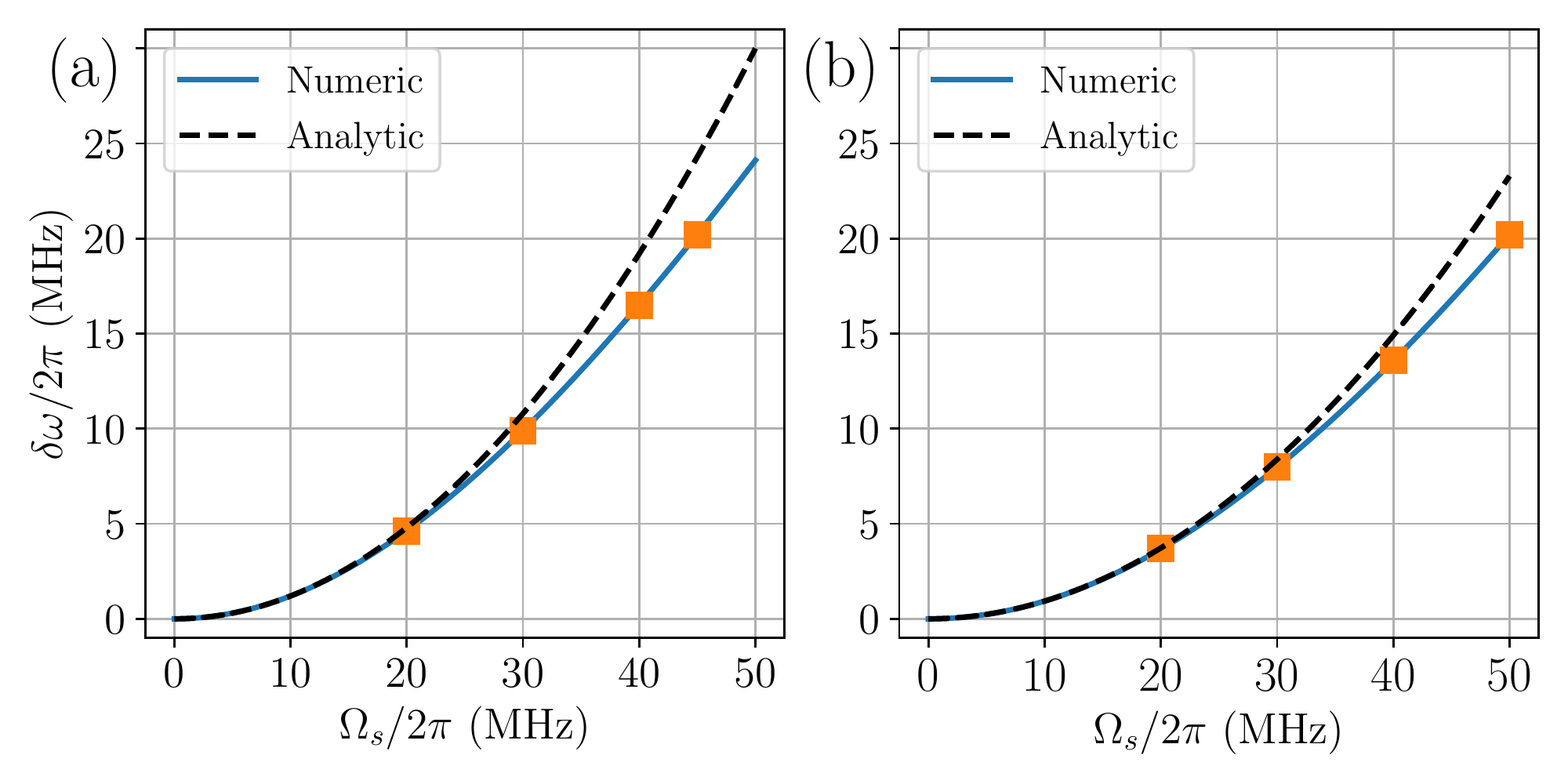}
\end{center}
\caption{The ac-stark shift versus the magnitude of the stark drive. (a) The detuning of
the stark drive applied to $Q_{1}$ is $\Delta_{s}/2\pi=50\,\rm MHz$, as marked
in Figs.~\ref{fig2}(a-c). (b) The detuning of the stark drive applied to $Q_{2}$
is $\Delta_{s}/2\pi=70\,\rm MHz$, as marked in Figs.~\ref{fig2}(d-f). Other
system parameters are the same as those used in Figs.~\ref{fig2}. The blue (light gray) solid lines
denote the results obtained by numerical calculation, and the black dashed lines for
analytical method, i.e., according to Eq.~(\ref{eq2}). The orange squares
mark the stark drive amplitudes which are used in Sec.~\ref{SecIIIB}.}
\label{fig3}
\end{figure}

According to the above discussion, to ensure high-fidelity two-qubit gate operations,
the stark drive frequency (i.e., detuning $\Delta_{s}$) should be chosen carefully to avoid inducing
parasitic interactions. Without a doubt, this will limit the available
range of the frequency and the amplitude of the stark drive. However, as shown
in Fig.~\ref{fig2}, there still exists available parameter regions, for which
one can avoid the detrimental parasitic interactions. As an example, the
black dashed lines in Figs.~\ref{fig2}(a) and Figs.~\ref{fig2}(d) denote
suitable frequencies ($\Delta_{s}/2\pi=\{50,70\}{\rm MHz}$) of the stark
drives applied to $Q_{1}$ and $Q_{2}$, respectively. Accordingly, Figures~\ref{fig3}(a)
and~\ref{fig3}(a) show the ac-stark shift $\delta\omega$ versus the stark drive
amplitude $\Omega_{s}$. The results shown in Fig~\ref{fig2} and Fig~\ref{fig3} indicate
that by varying the amplitude of the applied stark drive from 0 to $50\,\rm MHz$, the
ac-stark shift of the dressed qubit can be continually tuned from 0 to $20\,\rm MHz$
while avoiding detrimental parasitic interactions induced by the stark drives themselves.

In the following discussion, for illustration purposes, to examine the
performance of qubit control on the dressed qubits, the detuning $\Delta_{s}$
are chosen as $50\,\rm MHz$ and $70\,\rm MHz$ for the stark drives applied to
the $Q_{1}$ and $Q_{2}$, respectively. Additionally, we restrict our discussion
to the case where the stark drive is only applied to one of the two qubits. The
main reason will be given in Sec.~\ref{SecIV}.

\subsection{Qubit initialization and readout}\label{SecIIIA}

\begin{figure}[tbp]
\begin{center}
\includegraphics[keepaspectratio=true,width=\columnwidth]{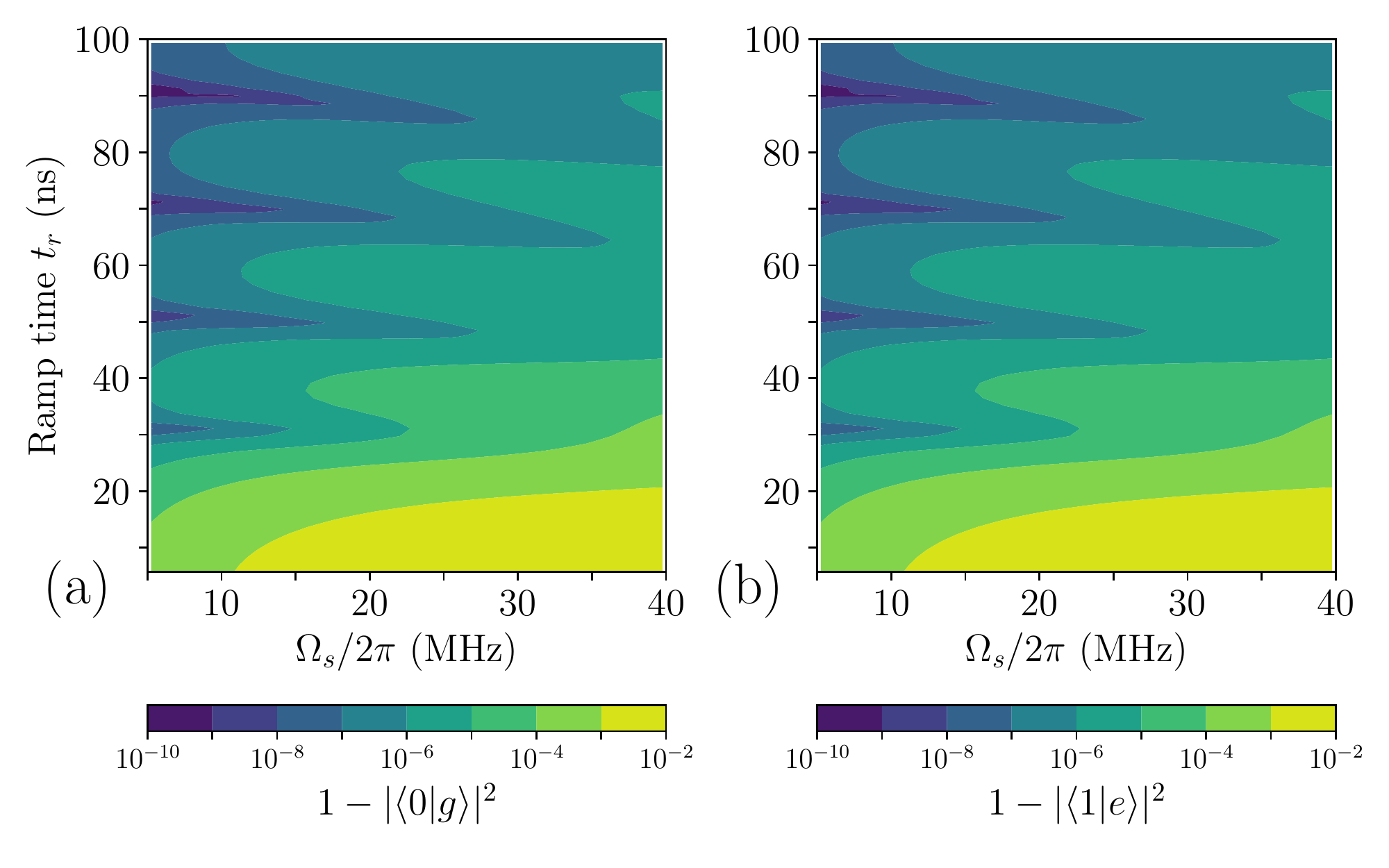}
\end{center}
\caption{Error for $Q_{1}$ in the adiabatical map from the dressed-qubit state $|0(1)\rangle$ to
the bare state $|g(e)\rangle$ as a function of the ramp time $t_{r}$ and the
stark drive amplitude $\Omega_{s}$. (a) Map from the dressed state $|0\rangle$ to the bare state $|g\rangle$.
(b) Map from the dressed state $|1\rangle$ to the bare state $|e\rangle$. Here, the
drive detuning is $\Delta_{1}/2\pi=50\,\rm MHz$
and pulse shape for the ramping process is $\Omega(t)=\Omega_{s}(1-\cos[\pi(t-t_{r})/t_{r}])/2$.
Other system parameters are the same as those in Fig.~\ref{fig2}.}
\label{fig4}
\end{figure}

In the presence of the always-on stark drive, the computational basis is the microwave-dressed
basis. Therefore, the qubit control, including qubit initialization and readout, should be also
operated on the dressed state. Similar to the procedure designed for dressed spin
qubits \cite{Seedhouse2021} or Floquet qubits \cite{Huang2021},
by slowly ramping up or down the amplitude of the stark drive, the bare qubit states $\{|g\rangle,\,|e\rangle\}$
and the dressed qubit states $\{|0\rangle,\,|1\rangle\}$ can be adiabatically
mapped to each other, thus enabling the initialization and readout of the
dressed qubit. In this way, the qubit initialization can be realized by firstly initializing the qubit in
the bare qubit state, e.g. $|g\rangle$, and then adiabatically mapping to
the corresponding dressed state, e.g. $|0\rangle$. While for implementing qubit readout, the reverse
procedure is firstly applied, thus mapping the dressed state to the corresponding bare state, and then
the traditional readout can be employed. Therefore, the fidelities of the
dressed-qubit initialization and readout depend on the applied ramp process. Figure~\ref{fig4} shows the
fidelity of the adiabatical map from the dressed-qubit state $|0(1)\rangle$ to the bare
state $|g(e)\rangle$ versus the ramp time $t_{r}$ and the stark drive
amplitude for $Q_{1}$ with the stark detuning $\Delta_{s}/2\pi=50\,\rm MHz$. The pulse shape for ramp down process
is $\Omega(t)=\Omega_{s}(1-\cos[\pi(t-t_{r})/t_{r}])/2$, where $t_{r}$ denote
the ramp time. As shown in Fig.~\ref{fig4}, high-fidelity ($99.9\%$) state maps can be realized
within 30 ns.

However, note that for readout, especially for the dressed state $|1\rangle$, after
mapping back to bare state $|e\rangle$ and without the protection of the stark drive, the readout fidelity
may be limited by the lifetime of state $|e\rangle$. To address this, one may take the excited state
promotion scheme, i.e, applying an additional $\pi$-pulse between the states $|1\rangle$ and
$|2\rangle$ before the adiabatic map and the measurement, as demonstrated in previous works \cite{Mallet2009,Elder2020,Jurcevic2021}.
This scheme may effectively extend the lifetime of the bare state $|e\rangle$, thus improving the readout fidelity.

\subsection{Gate operations}\label{SecIIIB}

In this subsection, we examine the implementation of single-qubit X gates and two-qubit CZ gates
on the proposed two-qubit system, in which a stark drive is applied to one of the
two qubits. Note that in the following discussion, in the two-qubit system, single-qubit gates for
one qubit are tuned up and characterized with the other qubit in its ground state.

\subsubsection{Single-qubit gate operation}\label{SecIIIB1}

\begin{figure}[tbp]
\begin{center}
\includegraphics[keepaspectratio=true,width=\columnwidth]{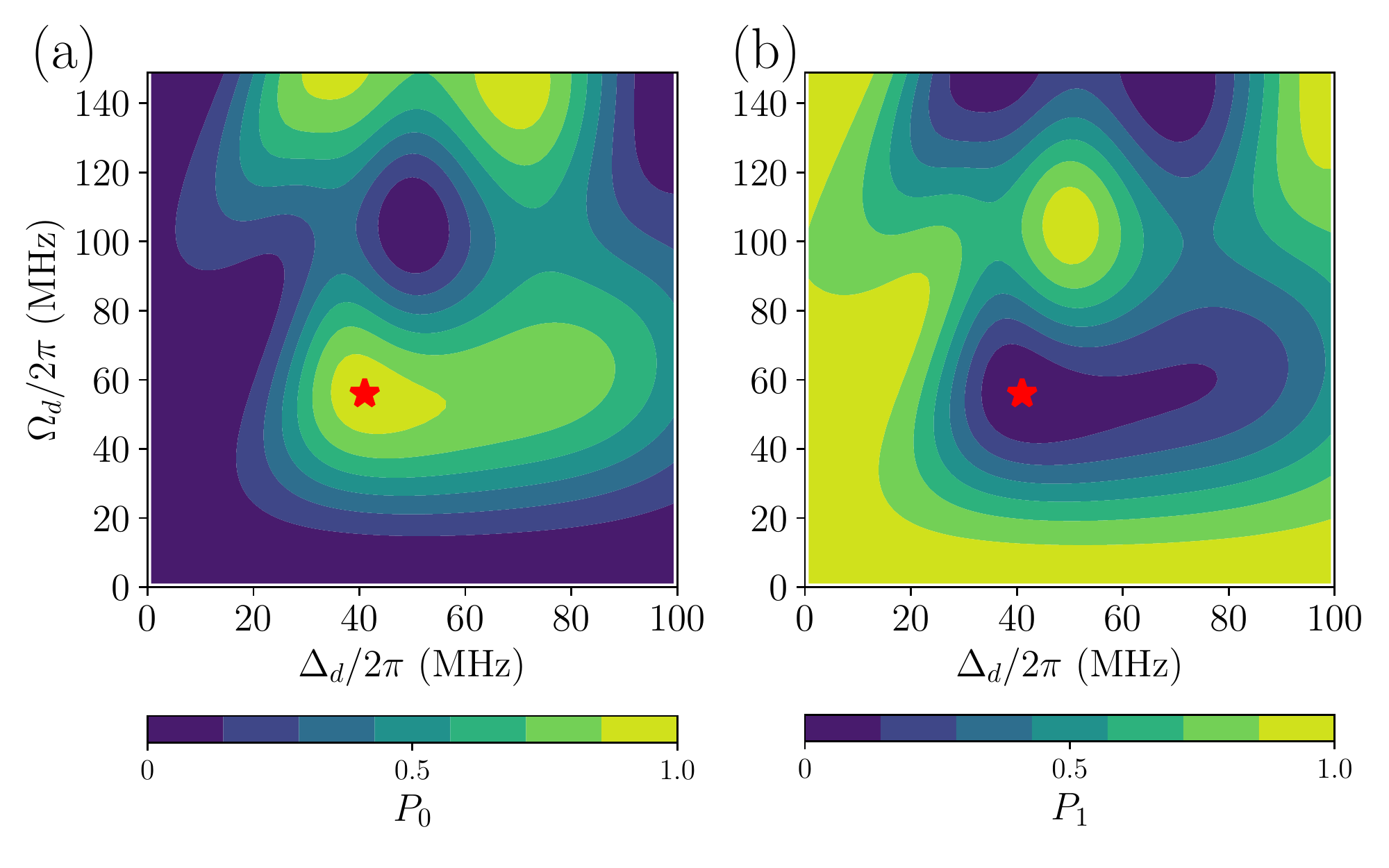}
\end{center}
\caption{Power Rabi oscillations for the microwave-dressed qubit. Population (a) $P_{0}$
and (b) $ P_{1}$ ($P_{i}$ denotes the population in the dressed state $|i\rangle$ at
the end of the gate operation with fixed gate length $t_{g}=20\,\rm ns$) as a function of
the driving amplitude $\Omega_{d}$ and the driving detuning $\Delta_{d}$ of the pulse drive
from the stark drive frequency. The red (light gray) star indicates the optimal parameter set for realizing
a single-qubit X gate. Here, $Q_{1}$ is initialized in the dressed state $|1\rangle$.
Other system parameters are the same as in Fig.~\ref{fig2}. Note here that due
to the weak anharmonicity of transmon qubits, during the gate operations, residual
populations in the non-computational states, such as the second excited state, can
exist. Thus, the sum of populations in computational states $|0\rangle$ and $|1\rangle$
is less than $1$. With the help of the DRAG scheme \cite{Motzoi2009}, here, the
population leakage is suppressed below $2\times10^{-5}$ over the entire region.}
\label{fig5}
\end{figure}

Here, we consider the implementation of single-qubit X gates.
During the single-qubit gate operations, the bus frequency is
biased at $5.7\,\rm GHz$, giving rise to a residual $ZZ$ coupling below $15\,\rm kHz$.
For both the microwave-dressed qubit and the bare qubit, single-qubit X
gates are realized using the derivative removal by
adiabatic gate (DRAG) scheme \cite{Motzoi2009}. The pulse shape is $\Omega(t)=\Omega_{x}(t)+i\Omega_{y}(t)$
with
\begin{eqnarray}
\begin{aligned}\label{eq9}
\Omega_{x}(t)=\frac{\Omega_{d}}{2}[1-\cos(2\pi t/t_{g})],
\Omega_{y}(t)=-\frac{\alpha}{\eta}\dot{\Omega}_{x}(t),
\end{aligned}
\end{eqnarray}
where $\Omega_{d}$ is the peak pulse amplitude, $t_{g}$ is the gate time, $\alpha$ is
a free parameter for mitigating the leakage to nonoccupational energy levels. Similar to
Ref.\cite{Zhao2021}, the X gate is tuned up by adjusting the driving amplitude $\Omega_{d}$
and the detuning $\Delta_{d}$ of the pulse drive from the stark drive frequency for
a fixed gate length ($t_{g}=20\,\rm ns$) and $\alpha=0.5$.
Here, note that the optimal parameters for realizing X/2-gates
can be found similarly by performing the power Rabi characterization with two
consecutive pulses, and the single-qubit rotations around the Y-axis
can be realized by controlling the phase of gate pulses, as indicated by Eq.~(\ref{eqa7}).

As high-fidelity single-qubit gates on the bare qubit have been
well-demonstrated \cite{Kjaergaard2020}, here, we especially focus on
the implementation of single-qubit gates for the microwave-dressed qubit.
Firstly, we show that by applying an additional pulse drive to
the dressed-qubit, Rabi oscillations between the dressed qubit states
can be realized, thus offering a way to realizing
single-qubit gates. Considering that a stark drive with the drive
detuning $\Delta_{s}/2\pi=50\,\rm MHz$ and the
amplitude $\Omega_{s}/2\pi=30\,\rm MHz$ is applied to $Q_{1}$, Figures~\ref{fig5} shows
the power rabi oscillation versus the driving detuning $\Delta_{d}$ for the dressed qubit $Q_{1}$
initialized in the dressed state $|1\rangle$. The
red (light gray) star in Fig.~\ref{fig5}(a) indicates an optimal parameter sets for performing an
almost perfect Rabi oscillation between $|0\rangle$ and $|1\rangle$.

Note that for microwave-dressed qubits, although Rabi
oscillations can indeed be induced by adding an additional drive, as shown
in Fig.~\ref{fig5}, the dynamics seem more complex than that
for undriven bare qubits (see Appendix~\ref{A} for
details). This can be partially manifested by frequency difference between the
optimal frequency of the pulse drive and the frequency of the microwave-dressed qubit.
As marked by the red stars in Fig.~\ref{fig5}, the detuning $\Delta_{d}$ of the
optimal pulse drive frequency from the stark drive is about $40\,\rm MHz$. However, according
to the ac-stark shift in Fig.~\ref{fig3}(a), the detuning of the microwave-dressed
qubit $Q_{1}$ from the stark drive frequency should be about $60\,\rm MHz$ (recalling that
$\Delta_{s}/2\pi=50\,\rm MHz$, and the ac-stark shift is about $\delta\omega/2\pi=10\,\rm MHz$
for $\Omega_{s}/2\pi=30\,\rm MHz$).

\begin{figure}[tbp]
\begin{center}
\includegraphics[keepaspectratio=true,width=\columnwidth]{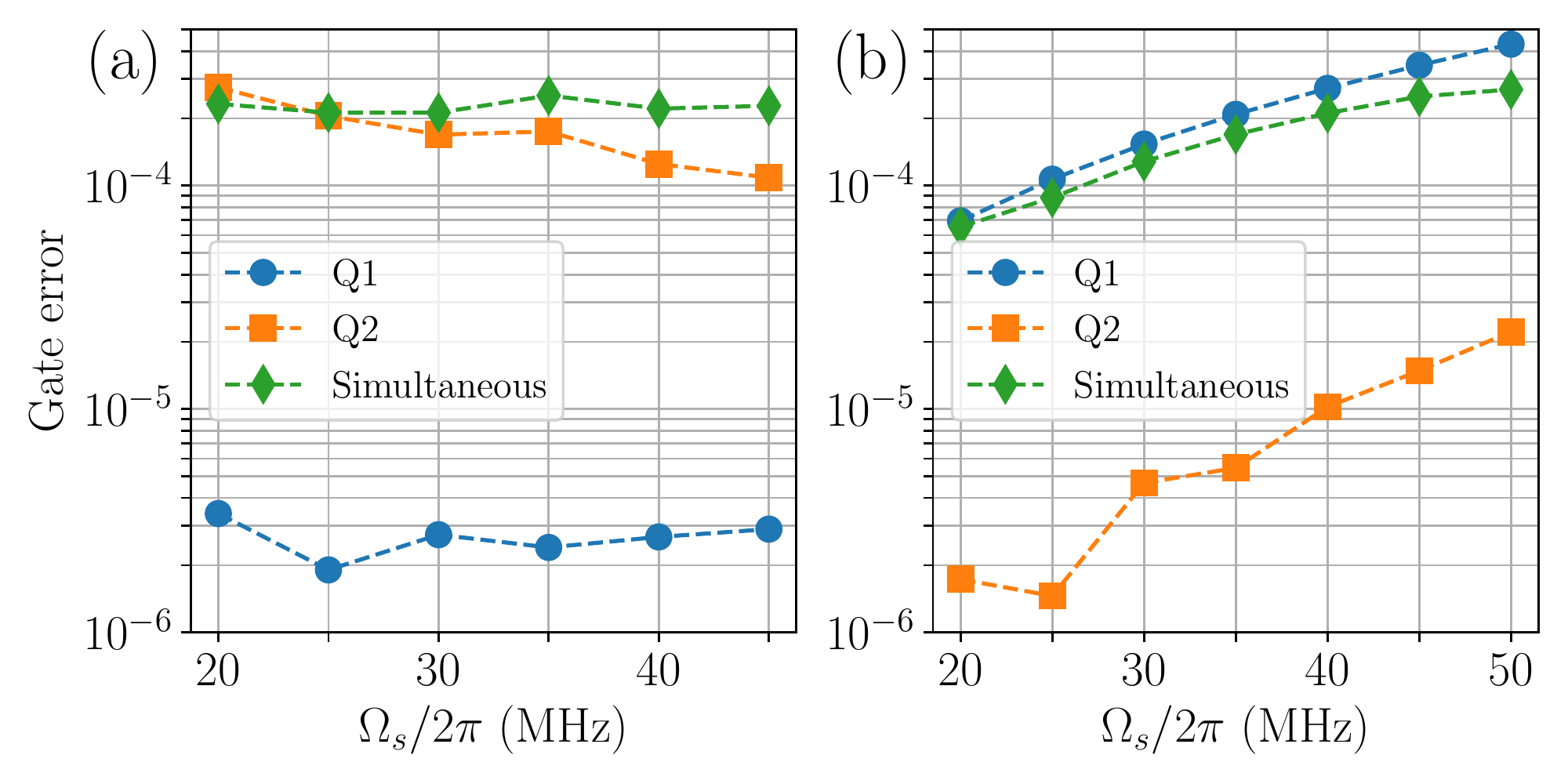}
\end{center}
\caption{Gate errors of the single-qubit X operations versus the stark drive
amplitudes $\Omega_{s}$. (a) The stark drive is applied to $Q_{1}$ with the drive
detuning $\Delta_{s}/2\pi=50\,\rm MHz$. (b) The stark drive is applied to $Q_{2}$ with the drive
detuning $\Delta_{s}/2\pi=70\,\rm MHz$. Other system parameters are the
same as in Fig.~\ref{fig2}.}
\label{fig6}
\end{figure}

After finding the optimal parameter sets for implementing X gates, we can
characterize the gate performance using the metric of gate fidelity \cite{Pedersen2007}.
Here, to quantify the performance of the (isolated) single-qubit X gate applied to
one qubit (i.e., target qubit) in the coupled qubit system, we always assume that the other
nearby qubit (i.e., spectator qubit) is in the ground state. Figure~\ref{fig6} shows the
gate performance of the implemented single-qubit gates. In addition, the gate fidelity for
simultaneously implementing single-qubit X gates is also presented. From these results, one can
conclude that in the presence of a stark drive applied to one of the two
coupled qubits, X gates applied to the bare qubit
show worse gate performances (gate error $\sim10^{-4}$) than that for the
microwave-dressed qubits (gate error $\sim10^{-5}$).
Meanwhile, the performances for the simultaneous
X gate operations are limited by the error from the
bare qubits. After examining the system dynamics during the gate operations,
these counterintuitive results can be explained by the interplay between the stark drive
and the cross-driving crosstalk due to the qubit-qubit
coupling \cite{Zhao2021}, i.e., drives, such as the stark drive or the
gate drive, applied to $Q_{1}$ can be felted by $Q_{2}$ and vice versa.

\begin{figure}[tbp]
\begin{center}
\includegraphics[keepaspectratio=true,width=\columnwidth]{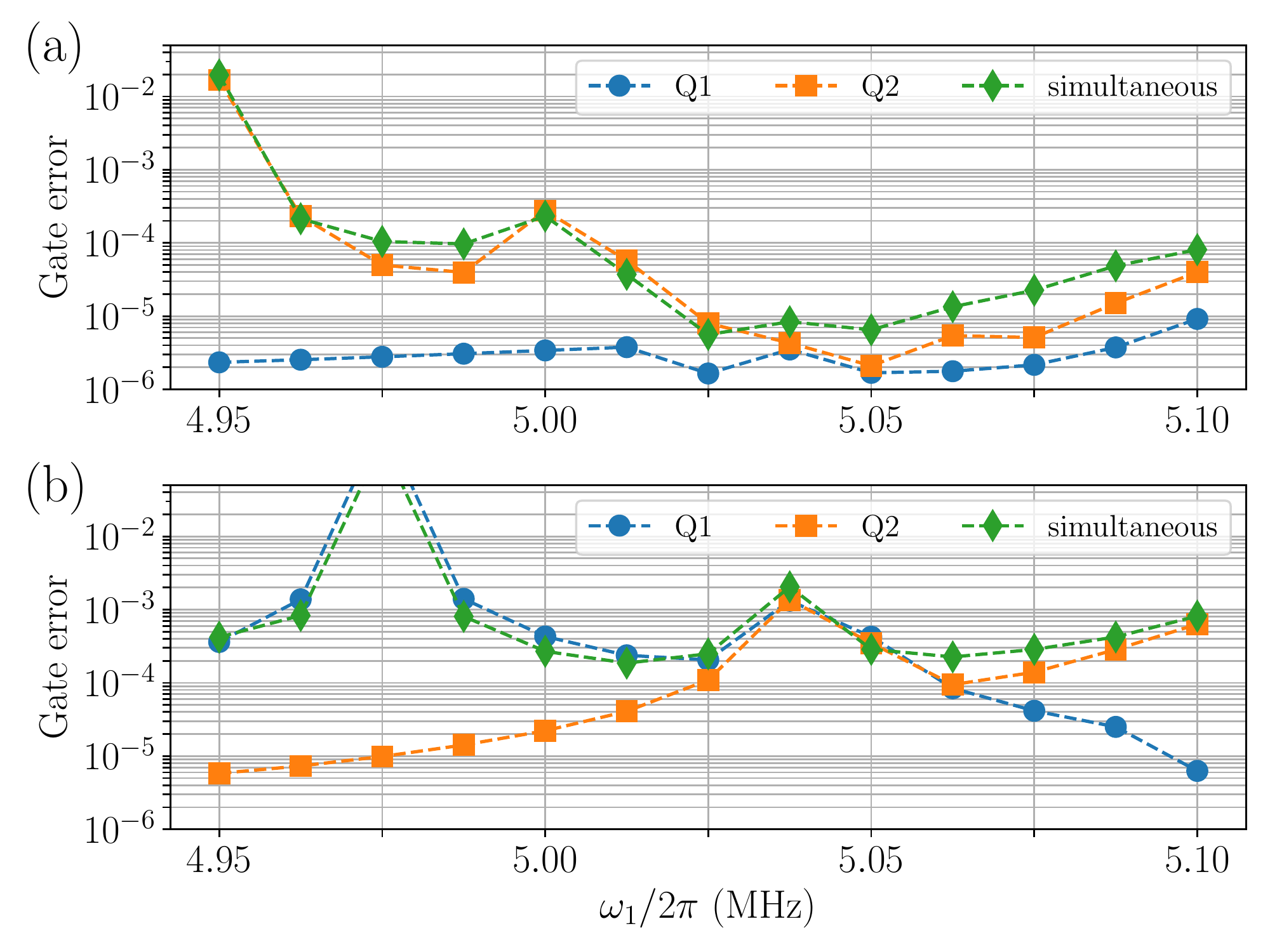}
\end{center}
\caption{Gate errors of the single-qubit X operations versus the frequency of $Q_{1}$
with $Q_{2}$ fixed at $5.2\,\rm GHz$. (a) The stark drive is applied to $Q_{1}$ with the drive
detuning $\Delta_{s}/2\pi=50\,\rm MHz$ and the drive amplitude $\Omega_{s}/2\pi=20\,\rm MHz$.
(b) The stark drive is applied to $Q_{2}$ with the drive
detuning $\Delta_{s}/2\pi=70\,\rm MHz$ and the drive amplitude $\Omega_{s}/2\pi=50\,\rm MHz$.
Other system parameters are the same as in Fig.~\ref{fig2}.}
\label{fig7}
\end{figure}

To further explore the exact nature of the extra error for bare qubits, in Fig.~\ref{fig7}, we show the
gate performance as a function of the frequency of $Q_{1}$ with $Q_{2}$ fixed
at $5.2\,\rm GHz$. Combined with inspections of the dynamics, we conclude that the
presence of the peaked gate error can be attributed to various parasitic resonance
transitions, resulting from the interplay between the stark drive and the
cross-driving crosstalk. For example, in Fig.~\ref{fig7}(a), where the stark drive
is applied to $Q_{1}$, the significant gate
error at $\omega_{1}/2\pi=5.0\,\rm GHz$ is caused by the three-photon
transition $|000\rangle\leftrightarrow|102\rangle$ and the two-photon
transition $|001\rangle\leftrightarrow|102\rangle$. In Fig.~\ref{fig7}(b), where the stark
drive is applied to $Q_{2}$, the significant gate
error at $\omega_{1}/2\pi=4.975\,\rm GHz$ results from the two-photon
transition $|000\rangle\leftrightarrow|002\rangle$ ($|100\rangle\leftrightarrow|102\rangle$), and
single-photon transition $|000\rangle\leftrightarrow|001\rangle$ ($|100\rangle\leftrightarrow|101\rangle$)
for $\omega_{1}/2\pi=5.0375\,\rm GHz$. Moreover, one can find that both the
target qubit and the spectator qubit are involved in the parasitic transitions.
Since under our definition of gate characterization,
we assume that the spectator qubit is always in its ground state, thereby, these parasitic
transitions can contribute to the isolated single-qubit errors, e.g., causing leakage
error.

Additionally, we note that for the target qubit $Q_{2}$, the peaked gate error
at $\omega_{1}/2\pi=4.95\,\rm GHz$, as shown in Fig.~\ref{fig7}(a), are caused by the
transitions $|000\rangle\leftrightarrow|002\rangle$ and $|001\rangle\leftrightarrow|002\rangle$.
The two transitions are solely the result of the cross-driving from
$Q_{1}$, which is driven by a stark tone with a frequency of $4.90\,\rm GHz$.
For $Q_{2}$, in principle, this cross-drive could be treated as an
unintended stark drive. However, since the frequency of this unintended drive is
on-resonance with the transitions involving high-energy levels, i.e,
$|1\rangle\leftrightarrow|2\rangle$ of $Q_{2}$, the descriptions that are based on
two-level approximation, such as in Appendix~\ref{A}, break down, and
the high-energy level $|2\rangle$ will affect the system dynamics
significantly. In this situation, we argue that this unintended cross-drive
will give rise to very nontrivial single-qubit addressing, causing the
extra error at $\omega_{1}/2\pi=4.95\,\rm GHz$ for $Q_{2}$.

Overall, the above analysis suggests that
although the presence of the stark drive has only almost negligible effects on
the gate performance for the microwave-dressed qubit, its effect on the gate
performance for the nearby coupled bare qubit, which is coupled to the
dressed qubit via a tunable bus, should be seriously considered. Recalling that
during the single-qubit gate operations, the residual $ZZ$ couplings are suppressed
below $15\,\rm kHz$. The above results further stress that:

(i) Besides residual $ZZ$
coupling, cross-driving crosstalk due to qubit-qubit transversal coupling should
also be taken into consideration for implementing high-fidelity gate
operations \cite{Zhao2021}. Here, the presence of the stark drive and the cross-driving
crosstalk together lead to parasitic transitions during single-qubit gate
operations on bare qubits.

(ii) The suppression of $ZZ$ coupling does not
always mean that the transversal coupling between qubits, which can induce the cross-driving
crosstalk, is also suppressed \cite{Zhao2021b}. Besides ensuring the suppression of the $ZZ$
coupling, these cross-driving induced parasitic transitions should also be
minimized. The results shown in Fig.~\ref{fig7}(b) suggest that for a given qubit-qubit
detuning, choosing a suitable stark frequency can avoid these parasitic resonance transitions.
In this way, the stark drive induced gate error is promising to be pushed
below $10^{-4}$.

\subsubsection{Two-qubit CZ gate operation}\label{SecIIIB2}

\begin{figure}[tbp]
\begin{center}
\includegraphics[keepaspectratio=true,width=\columnwidth]{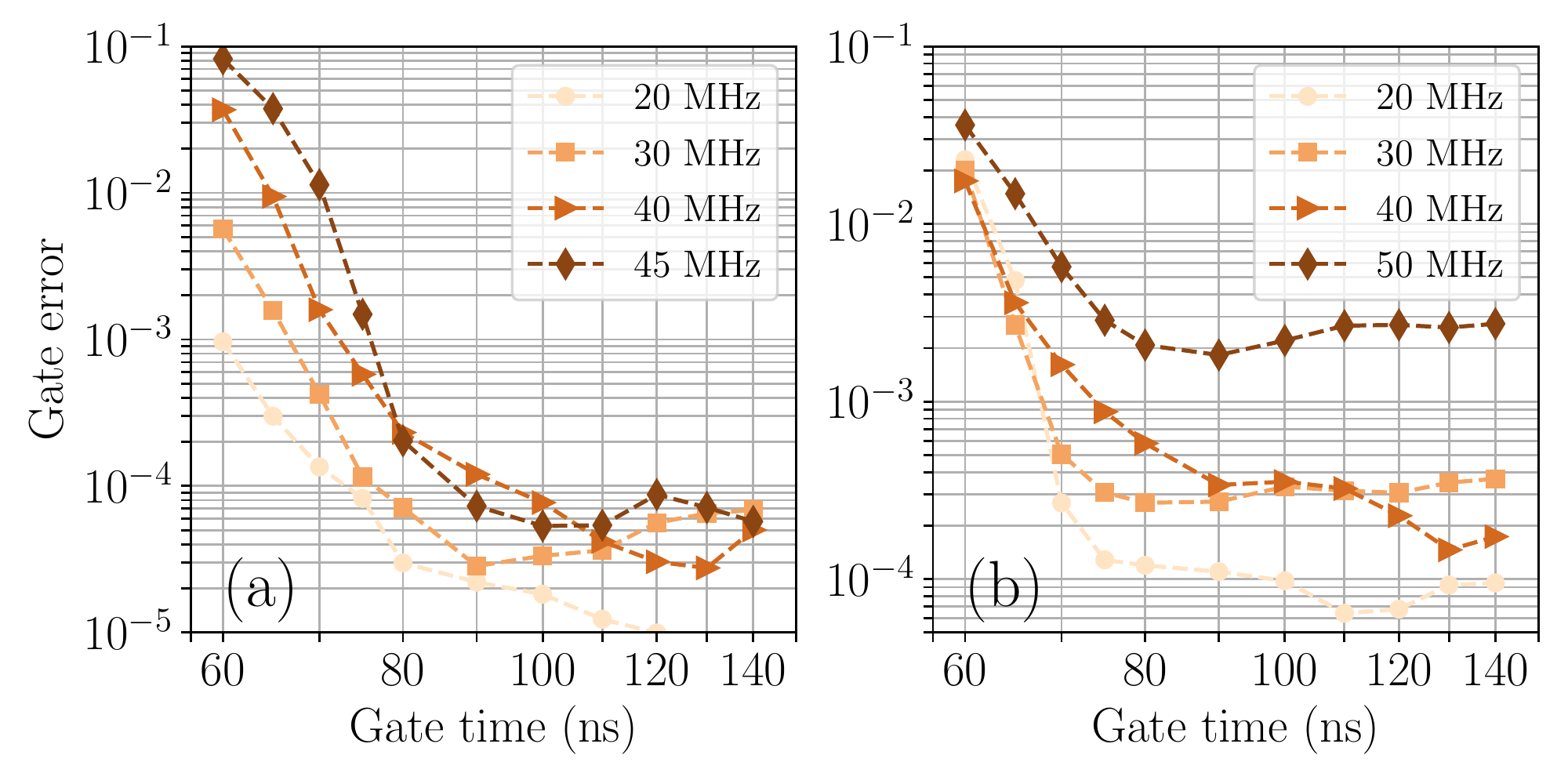}
\end{center}
\caption{Two-qubit CZ gate error versus the gate time with different stark drive
amplitudes. (a) The stark drive is applied to $Q_{1}$ with the drive
detuning $\Delta_{s}/2\pi=50\,\rm MHz$, and the drive amplitude
are $(20,\,30,\,40,\,45)\,\rm MHz$. (b) The stark drive is applied to $Q_{2}$ with the drive
detuning $\Delta_{s}/2\pi=70\,\rm MHz$, and the drive amplitude
are $(20,\,30,\,40,\,50)\,\rm MHz$. Other system parameters are the same as in Fig.~\ref{fig2}.}
\label{fig8}
\end{figure}

To implement two-qubit CZ gates on the two-qubit system, where a stark drive is
applied to one of the two qubits, we consider using the fast
adiabatic scheme \cite{Martinis2014,Zhao2021}. During the gate operations, the
tunable bus varies from the idle point at $5.7\,\rm GHz$ to the interaction
point and then comes back according to the fast adiabatic pulse (see Appendix~\ref{B} for
details). To implement fast CZ gates, e.g., sub-100-ns CZ gates, the interaction
point should be near the resonance point for the interaction $|101\rangle\leftrightarrow|020\rangle$,
where a large $ZZ$ coupling with the strength of $10\,\rm MHz$ exists, as shown in Fig.~\ref{fig2}.
Similar to Ref.\cite{Zhao2021}, the fidelity of the implemented
CZ gates are then obtained by optimizing the pulse shape. Figure ~\ref{fig8} shows the CZ gate
error for various gate times and different stark drive amplitudes.

In Fig.~\ref{fig8}(a), the stark drive is applied to $Q_{1}$, and
its detuning $\Delta_{s}$ is $50\,\rm MHz$. One can find that CZ gate errors
below $10^{-4}$ can be achieved with the gate time below 100 ns. More
strikingly, in general, the larger the stark drive amplitude $\Omega_{s}$, the
worse the CZ gate fidelity becomes. In addition, one can also find that for gate time greater
than certain values, gate error rises. Similar conclusions can also be
obtained for the case where the stark drive is applied to $Q_{2}$, as
shown in Fig.~\ref{fig8}(b). As the discussion given in Fig.~\ref{fig2},
these features can be explained by the presence of the parasitic
resonance interactions with tiny energy gaps. These parasitic resonance
interactions are caused by the stark drive, and involve both
the qubits and the bus. During the gate operations, the qubit system
will pass through or stay nearby to these parasitic resonance
points. Generally, for leakage error due to the
parasitic interactions with tiny strengths, the longer the
gate time, the larger the error becomes \cite{Zhao2021}. The opposite
is the case for leakage error resulting from the desired
interaction $|101\rangle\leftrightarrow|020\rangle$.
This trade-off results in the increased error
shown in Fig.~\ref{fig8} for gate time exceeding certain values.

\begin{figure}[tbp]
\begin{center}
\includegraphics[keepaspectratio=true,width=\columnwidth]{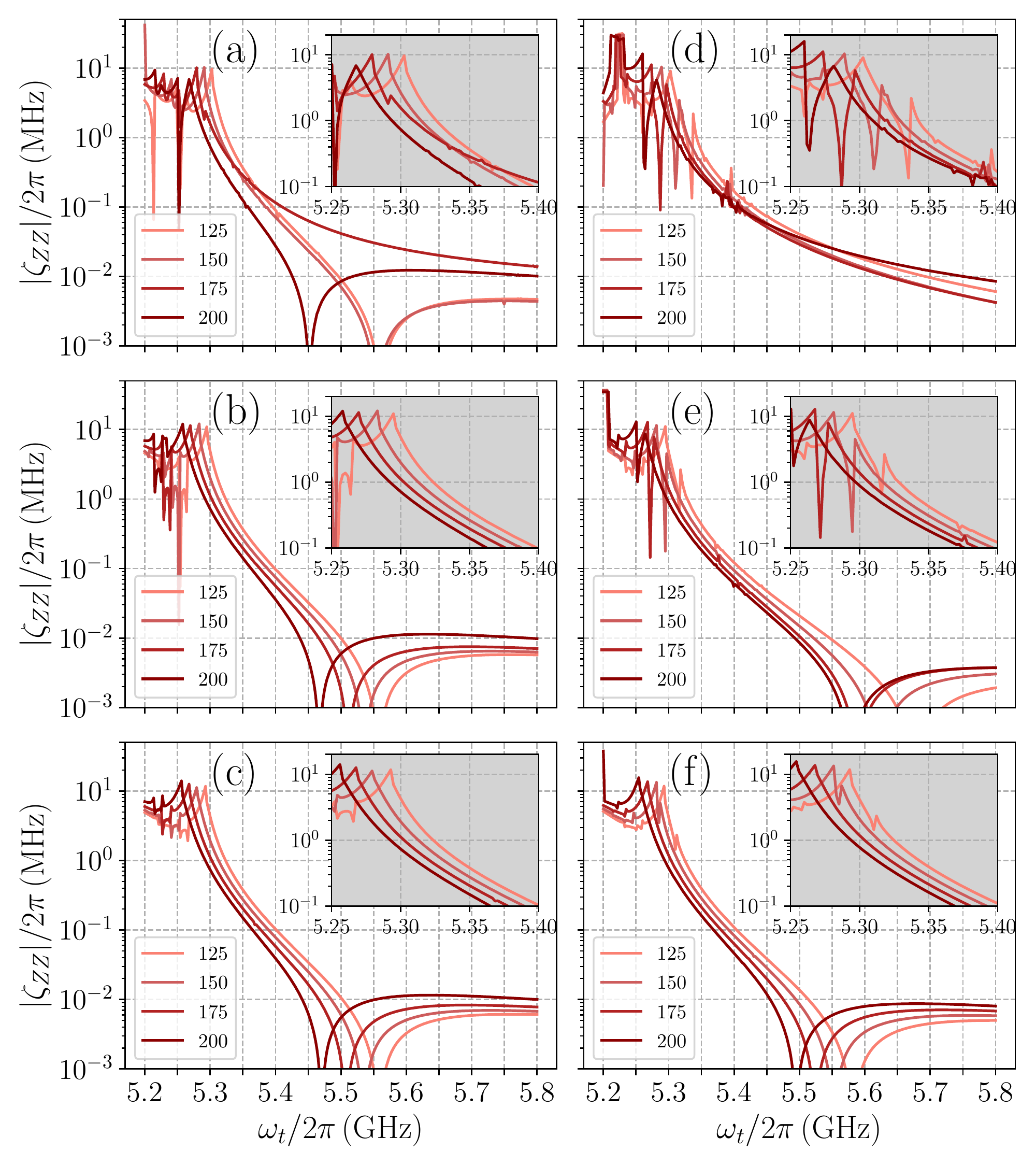}
\end{center}
\caption{$ZZ$ coupling strength $\zeta_{\rm ZZ}$ as a function of the bus frequency ($\omega_{t}$)
with different qubit-qubit detuning $(125,\,150,\,175,\,200)\,\rm MHz$. Here, $Q_{2}$ is
fixed at $5.2\,\rm GHz$. (a-c) The stark drive is applied to $Q_{1}$ with the
drive detuning $\Delta_{s}/2\pi=50\,\rm MHz $, and (d-f) for $Q_{2}$ with $\Delta_{s}/2\pi=70\,\rm MHz$.
The amplitude $\Omega_{s}$ of the stark drive are $|\Delta_{s}|$ in (a,d), $|\Delta_{s}|/2$ in (b,e),
and $|\Delta_{s}|/4$ in (c,f). The insets enlarge the curves nearby the resonance points of
the interaction $|101\rangle\leftrightarrow|020\rangle$. Other system parameters are the same
as in Fig.~\ref{fig2}.}
\label{fig9}
\end{figure}

To show explicitly the presence of the parasitic resonance interactions,
Figure~\ref{fig9} shows the $ZZ$ coupling strength versus the bus
frequency with different qubit detunings. Here, the $Q_{2}$ is
fixed at $5.2\,\rm GHz$. As shown in Figs.~\ref{fig9}(a-c), for a given qubit-qubit
detuning, when the amplitude of the stark drive applied to $Q_{1}$ decreases, dips and
peaks caused by the parasitic interactions also slowly
disappear, in line with the expectation (see also in Fig.~\ref{fig2}). Similar
results can also be obtained for the case where the stark drive applied to $Q_{2}$, as shown
in Figs.~\ref{fig9}(d-e). Moreover, compared with the case
where the stark drive applied to $Q_{1}$, there exist more dips and peaks
for the case where the stark drive applied to $Q_{2}$. This explains that the gate performance is worse
when the stark drive is applied to $Q_{2}$, as shown in Fig.~\ref{fig8}(b). This suggests that due to
these parasitic interactions, even without the consideration
of the qubit decoherence, long-time gates do not promise a better
gate performance.

In addition, from the results shown in Fig.~\ref{fig9}, we can conclude
that the qubit-qubit detuning should be designed carefully to suppress the
detrimental effect from the parasitic interactions in the tunable-bus architecture. As shown in
Fig.~\ref{fig9}(d-f), the smaller the qubit-qubit detuning, the
more the parasitic interactions exist. Overall, the result shown
in Fig.~\ref{fig8} and~\ref{fig9} suggests that besides the stark drive
frequency and amplitude, the qubit-qubit detuning can also act as a control knob
for avoiding the parasitic interactions. In this way, even with the
presence of the stark drive for inducing the qubit frequency shift up to $20\,\rm MHz$,
sub-100-ns CZ gates can still be realized in the tunable-bus architecture with
gate errors approaching $10^{-4}$.

\section{Feasibility}\label{SecIV}

As demonstrated in previous works \cite{Klimov2018,Carroll2021}, for the
state-of-the-art transmon qubits, the TLS defects that are coupled strongly
to qubits are generally sparse in spectrum. We thus expect
that tuning qubit frequency by few $\,\rm MHz$ to tens of $\,\rm MHz$
can mitigate the loss due to the dominant TLS defect, thus stabilizing
the qubit relaxation time. For fixed-frequency transmon qubits, this
can be realized by using the off-resonance stark drive induced
frequency shift. According to the discussion given in Sec.~\ref{SecIII}, we show
that although presence of stark drives can indeed cause additional gate
errors in the tunable-bus architecture utilizing fixed-frequency
transmon qubits, high-fidelity gate operations can still be achieved by
carefully choosing stark drive frequency, drive amplitude, and
qubit-qubit detuning. As shown in Figs.~\ref{fig3},~\ref{fig6}, and~\ref{fig8},
we illustrate that one can effectively tune the qubit frequency up
to $20\,\rm MHz$ through the ac-Stark shift, while implementing universal
gates with errors below $0.001$ on the tunable-bus architecture. By optimizing the
stark drive frequency and amplitude, even lower gate
error, e.g., below $0.0001$, should be achieved. We thus expect that
for fixed-frequency transmon qubits, the ac-stark shift should be a feasible
tool for mitigating TLS-induced performance fluctuations while keeping the
minimal impact on qubit control.

Note that the present study restricts to the case where the stark drive is
only applied to one of the two qubits. Although the present analysis can also
be applied to the case where both two qubits are subjected to the stark
drives. We expect that in that case, the stark drive induced parasitic couplings
will further limit the available parameter regions for implementing
high-fidelity gate operations. Furthermore, for the state-of-art transmon
qubit device, the TLS defects coupled strongly to qubits are generally
both few in numbers and sparse in the spectrum \cite{Klimov2018,Carroll2021}.
Thus, one can reasonably expect that the circumstance, where two coupled
fixed-frequency transmon qubits are both coupled strongly to TLS defects
at the same time, should be very rare.

\section{conclusion}\label{SecV}

In this work, we explore the possibility of combating TLS-induced temporal fluctuations
in relaxation rates of fixed-frequency qubits with microwave-dressed states.
Our analysis focus on the tunable-bus architecture, where fixed-frequency qubits are coupled
via a tunable bus, however, we expect that many of our conclusions may also
be applied to other qubit architectures utilizing fixed-frequency transmon qubits.
While during gate operations, the stark drive can lead to
additional gate errors due to parasitic interactions induced by the
stark drive, one can mitigate their detrimental impacts on qubit control by
carefully choosing the drive parameters and the system parameters. In
this way, we show that one can effectively tune the qubit frequency through ac-Stark
shift up to $20 \,\rm MHz$ while keeping minimal impacts on the qubit control
including qubit initialization, qubit readout, and gate operations.

\begin{acknowledgments}

We acknowledge helpful discussions with Yanwu Gu and Xinsheng Tan. This work
was supported by the Beijing Natural Science Foundation (Grant No.Z190012), the National
Natural Science Foundation of China (Grants No.11890704, No.12004042), and the
Key-Area Research and Development Program of Guang Dong Province (Grant No. 2018B030326001).
T.M. was supported by the National Natural Science Foundation of China (Grant No.11905100).

\end{acknowledgments}

\appendix

\section{single-qubit gate operations with microwave-dressed qubits}\label{A}

For illustration purpose and without loss of generality, here we begin our analysis based on a
two-level system subjected to two microwave drives, for which its dynamics is governed by the following Hamiltonian
\begin{equation}
\begin{aligned}\label{eqa1}
H_{\rm lab}=\frac{\omega_{q}}{2}\sigma_{z}+\Omega_{s}\cos(\omega_{s}t)\sigma_{x}+\Omega_{d}\cos(\omega_{d} t+\phi)\sigma_{x},
\end{aligned}
\end{equation}
where $\omega_{q}$ denotes the bare qubit transition frequency, $\Omega_{s}$ is the frequency of the stark drive (for effectively tuning the qubit frequency) with amplitude $\Omega_{s}$, and $\omega_{d}$ is the frequency of the gate pulse drive (for implementing single-qubit gate
operations) with amplitude $\Omega_{d}$. $\phi$ denotes the relative phase between the stark drive and the gate pulse drive.

After applying the RWA, and moving into the rotating
frame with respect to the off-resonant stark drive, the Hamiltonian
can be expressed as $H_{R}=H_{0}+H_{I}$ with
\begin{equation}
\begin{aligned}\label{eqa2}
&H_{0}=\frac{\Delta_{s}}{2}\sigma_{z}+\frac{\Omega_{s}}{2}\sigma_{x},
\\&H_{I}=\frac{\Omega_{d}}{2}[\sigma^{+}e^{-i(\Delta_{d} t+\phi)}+\sigma^{-}e^{+i(\Delta_{d} t+\phi)}],
\end{aligned}
\end{equation}
where $\Delta_{s}=\omega_{q}-\omega_{s}$ ($\Delta_{d}=\omega_{d}-\omega_{s}$) denotes the detuning of the qubit bare
frequency (pulse driving frequency) from the stark driving frequency.
Considering the unitary transformation $U_{1}={\rm exp}(-i\theta\sigma_{y}/2)$ with $\theta=\arctan(\Omega_{s}/\Delta_{s})$
that intends to diagonalizing $H_{0}$ in the dressed qubit basis, (i.e., the dressed basis is defined as the eigenstates of $H_{0}$)
\begin{equation}
\begin{aligned}\label{eqa3}
\{&|1\rangle\equiv\sin\frac{\theta}{2}|g\rangle+\cos\frac{\theta}{2}|e\rangle,
\\&|0\rangle\equiv\cos\frac{\theta}{2}|g\rangle-\sin\frac{\theta}{2}|e\rangle\},
\end{aligned}
\end{equation}
one can obtain the following dressed system Hamiltonian
$H_{\rm dress}=U_{1}^{\dagger}H_{R}U_{1}$, (i.e., the Hamiltonian in the dressed basis)
\begin{equation}
\begin{aligned}\label{eqa4}
H_{\rm dress}=&[\Delta-\Omega_{d}\sin\theta\cos(\Delta_{d} t+\phi)]\frac{Z}{2}
\\&+\Omega_{d}\cos\theta\cos(\Delta_{d} t+\phi)\frac{X}{2}
\\&+\Omega_{d}\sin(\Delta_{d} t+\phi)\frac{Y}{2},
\end{aligned}
\end{equation}
where $\Delta=\sqrt{\Delta_{s}^{2}+\Omega_{s}^{2}}$ denotes the microwave-dressed qubit detuning,
and $\{X=\cos{\theta}\sigma_{x}-\sin{\theta}\sigma_{z}, Y=\sigma_{y}, Z=\cos{\theta}\sigma_{z}+\sin{\theta}\sigma_{x}\}$
represent the Pauli operators defined on the dressed basis. According to the following unitary
transformation \cite{Oliver2005,Saiko2006,Tuorila2010}
\begin{equation}
\begin{aligned}\label{eqa5}
U_{2}={\rm exp}\left(-i\frac{Z}{2}\left[\Delta t-\frac{\Omega_{d}\sin\theta}{\Delta_{d}}\sin(\Delta_{d} t+\phi)\right]\right),
\end{aligned}
\end{equation}
and using the Jacobi-Anger relations, one can obtain the effective
Hamiltonian $H_{\rm J}=U_{2}^{\dagger}H_{\rm dress}U_{2}+i\partial_{t}(U_{2}^{\dag})U_{2}$, i.e.,

\begin{equation}
\begin{aligned}\label{eqa6}
H_{\rm J}=&\frac{\Omega_{d}\cos\theta}{4}(e^{i\Delta_{d} t+i\phi}+e^{-i\Delta_{d} t-i\phi})
\\&\times\left[e^{i\Delta t}S^{+}\sum_{n=-\infty}^{\infty}J_{n}(\frac{\Omega_{d}\sin\theta}{\Delta_{d}})e^{-in(\Delta_{d} t+\phi)}+h.c.\right]
\\&-\frac{\Omega_{d}}{4}(e^{i\Delta_{d} t+i\phi}-e^{-i\Delta_{d} t-i\phi})
\\&\times\left[e^{i\Delta t}S^{+}\sum_{n=-\infty}^{\infty}J_{n}(\frac{\Omega_{d}\sin\theta}{\Delta_{d}})e^{-in(\Delta_{d} t+\phi)}-h.c.\right],
\end{aligned}
\end{equation}
where $h.c.$ denotes the Hermitian conjugate, $J_{n}$ is the $n$th order Bessel function of the first kind, and $S^{\pm}=(X\pm iY)/2$.
Applying the RWA and dropping high-order Bessel functions, one can obtain the following effective Hamiltonian describing the
usual single-qubit driven terms,

\begin{equation}
\begin{aligned}\label{eqa7}
H_{\rm eff}&=\frac{\Omega_{d}\cos^{2}\frac{\theta}{2}}{2}
J_{0}(\frac{\Omega_{d}\sin\theta}{\Delta_{d}})\left[e^{-i\phi}S^{+}e^{i(\Delta-\Delta_{d})}+h.c.\right].
\end{aligned}
\end{equation}

For system parameters used in the present work, one has $|\Delta_{d}|\sim|\Delta_{s}|\sim|\Delta|$ and the
peak amplitude of the pulse drive $\Omega_{d}\sim|\Delta|$. We thus expect
that the RWA breaks and other terms given in Eq.~\ref{eqa6}, which involve high-order
Bessel functions, can have non-negligible effects on the system dynamics. When
considering $\Delta_{d}\approx\Delta$, these terms contribute as off-resonance
transitions during the single-gate operations, thus shifting the frequency of
the dressed qubit \cite{Saiko2006,Tuorila2010,Yan2017}. This can explain
the frequency mismatch discussed in Sec.~\ref{SecIIIB1}.

\section{Pulse shaping for implementing two-qubit CZ gates}\label{B}

In the present work, similar to Ref.\cite{Zhao2021}, we use the fast adiabatic
gate scheme for implementing CZ gates by tuning the bus frequency.
Here, for easy reference, we give brief descriptions
of the pulse shape for realizing CZ gates. During the CZ gate operation, the tunable bus
frequency $\omega_{t}$ varies from the idle point ($\theta_{i}$) to the
interaction point ($\theta_{f}$) and then back according to a fast adiabatic pulse.
Expressed in terms of Fourier basis functions, the pulse shape is described
as \cite{Martinis2014}

\begin{eqnarray}
\begin{aligned}\label{eqa8}
\theta(t)=\theta_{i}+ \frac{\theta_{f}-\theta_{i}}{2}\sum\limits_{n=1,2,3...}\lambda_{n}\left[1-\cos\frac{2n\pi t}{T}\right]
\end{aligned}
\end{eqnarray}
with constraints on the odd coefficients $\Sigma_{n\,\rm{odd}}\,\,\lambda_{n}=1$. Here, the control
angle is defined as $\theta\equiv\arctan(2J_{101}/\Delta_{101})$, where $J_{101}$ represents the
strength of the interaction $|101\rangle\leftrightarrow|020\rangle$, $\Delta_{101}$
represents the detuning of the qubit system from the resonance point of the
interaction $|101\rangle\leftrightarrow|020\rangle$, and $T$ is the gate time.

For implementing CZ gates, we consider the pulse shape defined in Eq.~\ref{eqa8} with
three Fourier terms, for which the associated coefficients are $\{\lambda_{1},\lambda_{2},1-\lambda_{1}\}$.
The free parameters $\{\lambda_{1},\lambda_{2},\theta_{f}\}$ are then determined
by numerically optimizing the CZ gate fidelity.

\end{document}